\documentstyle[aps,preprint,tighten,epsf,floats]{revtex}
\newcommand{\spartial}{\hspace{-1.2mm}\not\hspace{-.7mm}\partial}

\newcommand{\ben}{\begin{displaymath}}
\newcommand{\een}{\end{displaymath}}
\newcommand{\be}{\begin{equation}}
\newcommand{\ee}{\end{equation}}
\newcommand{\bea}{\begin{eqnarray}}
\newcommand{\eea}{\end{eqnarray}}
\newcommand{\eqn}[1]{\label{#1}}
\newcommand{\eq}[1]{Eq.~(\ref{#1})}
\newcommand{\eqs}[1]{Eqs.\ (\ref{#1})}
\newcommand{\fign}[1]{\label{#1}}
\newcommand{\fig}[1]{Fig.\ \ref{#1}}

\newcommand{\dG}{\delta {G}}
\newcommand{\dK}{\delta {K}}
\newcommand{\GG}{G}
\newcommand{\K}{K}

\begin{document}

\title{Pionic dressing of baryons in chiral quark models}
\author{Alexander N. Kvinikhidze\footnote{On leave from Mathematical 
Institute of Georgian Academy of Sciences, Tbilisi, Georgia.}, Michael 
C. Birse} 
\address{Theoretical Physics Group, Department of Physics and Astronomy, 
University of Manchester, Manchester M13 9PL, United Kingdom}
\author{Boris Blankleider}
\address{Department of Physics, The Flinders University of South 
Australia, Bedford Park, SA 5042, Australia}
\date{\today}
\maketitle

\begin{abstract}

We present a method for constructing the complete set of meson-exchange 
corrections to baryon observables described by covariant chiral quark 
models. The  meson corrections are expressed in terms of an unperturbed 
valence-quark Green's function. This method is illustrated for the case 
of an NJL model. It also enables us to discuss, in terms of 
approximations used for this unperturbed Green's function, the 
treatments of meson
 corrections in two other chiral quark models. 
\end{abstract}

\bigskip

\section{Introduction}

Chiral symmetry is one of the most important features of QCD at low energies. 
This approximate symmetry is spontaneously broken by the quark condensate,
leading to constituent masses for the quarks and to pions which are, to a good 
approximation, the corresponding Goldstone bosons. As a result pions are much 
lighter than all other mesons and so the cloud of virtual pions surrounding a 
baryon can make significant contributions to the properties of that baryon. Any 
realistic model for the structure of baryons should therefore take account of 
their pion clouds.

As approximate Goldstone bosons, pions with low momenta interact relatively
weakly with other hadrons. This forms the basis for chiral perturbation theory
(ChPT), in which hadron properties and scattering amplitudes are expanded in
powers of momenta and the pion mass (for a review, see\ \cite{BKM}).  It also
provides a motivation for treating the pion cloud perturbatively in chiral quark
models. Such a treatment ensures that at least the longest-range effects of the
pion cloud are described correctly, since these involve virtual pions with low
momenta. These effects can be checked by comparing contributions to hadron
properties that depend nonanalytically on the square of the pion mass with the
model-independent predictions of ChPT.

A range of chiral quark models for baryon structure can be found in the
literature.  These include chiral or ``cloudy" bag and soliton
models\cite{thomas,birse} and chiral versions of constituent quark
models\cite{CCQM,fern,GR,GPPVW}. In all of these, pions are introduced as
elementary degrees of freedom and in many cases their contributions are
calculated perturbatively.

In contrast, models of the Nambu--Jona-Lasinio (NJL) type are based on
relativistic quarks only, and the quark condensate and pions are generated by
the interactions between these quarks (for a review, see:\ \cite{klev}).  As
such they represent a step closer to QCD, although problems with implementing
confinement remain to be solved satisfactorily. In these models baryon states
have been constructed by solving the relativistic Faddeev
equation\cite{buck,IBY,huang}, but this approach does not include the pion
cloud.\footnote{We should mention that a completely different approach to
baryons in these models uses a mean-field treatment where baryons are described
as Skyrmion-like solitons\cite{alkofer,christov}. This is expected to be valid
in the limit of a large number of colours. However in the realistic case of
three colours the Faddeev approach is likely to be more appropriate. An attempt
to hybridise the two approaches can be found in Ref.\ \cite{zueck}.} The part of
the cloud corresponding to the pairwise meson exchange between quarks has been
studied in Ref.\ \cite{ishii}. Recently a covariant version of the treatment 
used in the cloudy-bag model has been applied to calculate pion cloud effects 
based on a Faddeev approach to the nucleon\cite{HORSTT,OT}. Also, a quite 
different, noncovariant approach to the calculation of pion cloud effects in 
models with composite pions has been proposed by Bicudo {\it et al.} \cite{BKR}.

In the present work we address the problem of mesonic corrections within an NJL
model, deriving covariant expressions for the complete set of one-loop
corrections to baryon masses and couplings to currents. The approach is based on
the ``gauging of equations" method \cite{nnn} for constructing photon couplings
to strongly interacting systems of particles. Our approach is applicable to
any set of one-boson contributions to such systems and was recently used to
construct a complete set of electromagnetic corrections\cite{Adel98}. As has 
been noted previously in the context of NJL models for 
mesons\cite{DSTL,NBCRG,oertel,PB}, such completeness is important since it ensures 
that physical quantities calculated in this approach satisfy the relevant symmetry
constraints. The general features of the approach also allow us to comment on
the approximations used in cloudy-bag and chiral quark models, which have
recently been the subject of some debate\cite{TK1,TK2,gloz1,gloz2}.

\section{Covariant mesonic corrections}

A method has recently been presented for calculating all possible 
electromagnetic corrections, at order $e^2$, to any quark or hadronic 
model whose strong interactions can be represented nonperturbatively by 
integral equations \cite{Adel98}. We will use this method to calculate in 
a covariant manner all mesonic corrections to a relativistic quark model. 
For this purpose all we need to do is to summarise the main 
model-independent results of Ref.\ \cite{Adel98}, applied to exchange of 
mesons rather than photons.

In such a model, the Green's function $G$ describing a bare system of
quarks satisfies an integral equation whose symbolic form is
\be
G=G_0+G_0 K G,       \eqn{G}
\ee
where $G_0$ is the ``free" Green's function. Note that this need not 
describe the propagation of three free quarks. In a Faddeev approach, for 
instance, it is the Green's function for one free quark and an interacting
pair of quarks (often referred to as a diquark, although the pair need 
not be bound).

The lowest-order mesonic correction to $G$ is denoted by $\dG$. This should 
contain all possible insertions of the meson propagator, $D_{\mu\nu}$. 
In chiral quark models with elementary mesons this is just a standard 
boson propagator, but in NJL-type models it is essentially
the quark-antiquark $T$-matrix, as discussed below. A four-vector-like 
notation is used for the indices $\mu$ and $\nu$ which label the meson 
channels. This will allow for applications to models with vector or axial 
mesons. The propagator is diagonal in its indices, except for the mixing 
between pseudoscalar and axial channels which occurs in models with spin-1 
mesons. The set of corrections includes mesonic insertions within the kernel 
$K$ which do not start or finish on an external quark line. These
provide a contribution $\dK$ to the kernel. Similarly insertions within
the free Green's function $G_0$ provide a contribution $\dG_0$.

All contributions with a meson which connects two elements of the
integral equation \eq{G} are constructed by the ``gauging of 
equations" method \cite{nnn} which in this case attaches an external
meson in all possible ways to each of the unperturbed quantities 
$G_0$, $K$ and $G$. The corresponding vertices are given by
\be
\Gamma_0^\mu = \GG_0^{-1} \GG_0^\mu \GG_0^{-1}, \hspace{1cm}
\Gamma^\mu = \Gamma_0^\mu + \K^\mu,\hspace{1cm}
\GG^\mu = \GG \Gamma^\mu\GG. 
\eqn{G^mu}
\ee
An equation for the complete set of lowest-order mesonic corrections 
to $G$, denoted by $\dG$, then follows from \eq{G}:
\be
\dG=\dG_0+\dG_0KG+G_0\dK G+G_0K\dG
+\left(G^\mu_0 K^\nu G+
G^\mu_0K G^\nu+G_0K^\mu G^\nu\right)D_{\mu\nu}.  \eqn{dG}
\ee
Writing
\be
\dG = \GG\Delta\GG,
\ee
this can be formally solved to give
\be
\Delta= \dK + \GG_0^{-1}\dG_0\GG_0^{-1}
+\left(\Gamma^\mu\GG\Gamma^\nu-\Gamma_0^\mu \GG_0\Gamma_0^\nu\right)
D_{\mu\nu}.
\eqn{Delta}
\ee
The quantity $\Delta$ consists of the complete set of one-loop mesonic corrections 
to the unperturbed kernel $K$. The third term on 
the right hand side (RHS) contains
the physical intermediate
 states of three quarks plus a meson. There are 
significant cancellations between 
the second and fourth terms, leaving 
contributions which correct the $G_0$ piece of
 the third term for omissions and 
overcounting.
 Note that the contributions of these terms include disconnected
pieces where the particles in $G_0$ are individually dressed by meson loops. These
can be easily recast as corrections to the ``free'' propagator $G_0$.

Although our main focus here is on covariant approaches, it is worth noting that
the same formalism also applies to time-ordered perturbation theory. The only 
difference is that the Green's functions involved are functions of three-momenta 
rather than 
four-momenta.
 A method for applying time-ordered perturbation theory 
to mesonic corrections in models of this type has recently been outlined in 
Ref.~\cite{BKR}, although that work proposed truncating the sum over intermediate 
baryon states to just the nucleon and $\Delta$. However, within a time-ordered 
approach, it is difficult to ensure that vertex functions satisfy the constraints of
Lorentz invariance. This is necessary if one wants to calculate processes involving
finite momentum transfer to a baryon, or to treat correctly recoil of the 
intermediate baryons in meson-loop diagrams. In addition, a time-ordered approach
would make it difficult to include quark ``z-diagrams", 
since these would go beyond a basis of states with three valence quarks (plus 
mesons). Such effects are expected to be important in models where the quarks 
have strong couplings to scalar mesons and couplings to pions with a pseudoscalar 
form. It is also difficult to regularise time-ordered approaches to NJL models in
a way that respects Lorentz invariance. For these reasons we have concentrated 
here on developing a covariant treatment.

\section{NJL model for the nucleon}

The NJL model\cite{klev} is a relativistic quark model which reflects many  
aspects of the spontaneously broken chiral symmetry of QCD. It
therefore provides a useful framework within which to investigate the
role of mesonic corrections.

The simplest version of the model is defined in terms of an
isospin-doublet colour-triplet quark field $\psi$ and has
the Lagrangian density 
\be
{\cal L}=\bar\psi\left(i\spartial-m_0\right)\psi+g_\pi\left[(\bar\psi\psi)^2-
(\bar\psi\gamma_5\mbox{\boldmath $\tau$}\psi)^2\right]  \eqn{NJLL}
\ee
where {\boldmath $\tau$} is the vector of isospin Pauli matrices, and
$m_0$ is the current quark mass (which explicitly breaks chiral symmetry).

Chiral symmetry is spontaneously broken in this model by condensation of
quark-antiquark pairs. These generate a constituent mass for the quarks,
which can be found by solving the Schwinger-Dyson (SD) equation  for the 
quark propagator
\be
S(p)=S_0(p)+S_0(p)\Sigma(p)S(p)    \eqn{DS}
\ee
where $S_0(p)$ is the bare quark propagator and the self-energy $\Sigma$
is, in the Hartree approximation,
\be
\Sigma(p)=-ig_\pi\int \frac{d^4k}{(2\pi)^4}\mbox{tr}[S(k)].
\eqn{Hartree}
\ee
Here ``tr'' denotes a trace over Dirac, colour and flavour indices. The 
integral over four-momenta is, of course, divergent and so needs to be
regularised. This may be done either by using one of the various cut-off
schemes which can be found in the literature, or by introducing form-factors
to smear out the local interaction. Provided that the coupling $g_\pi$ is 
large enough, this SDE has a nonzero solution for $\Sigma$.

The quark-antiquark $T$-matrix can be found by solving a Bethe-Salpeter (BS)
equation, which in the random-phase approximation is a simple algebraic
equation. For example, in the pion channel this yields
\be
D_\pi(p)=\frac{4ig_\pi}{1-2ig_\pi\Pi_\pi(p^2)}, \eqn{Dp}
\ee
where the basic quark-antiquark loop integral is
\be
\Pi_\pi(p^2)\delta_{ij}=\int
\frac{d^4k}{(2\pi)^4}\mbox{tr}\left[i\gamma_5\tau_i S(p+k)i\gamma_5\tau_j
S(k)\right]. \eqn{Pip}
\ee
In the simplest NJL model there is a similar $T$-matrix in the scalar isoscalar
($\sigma$-meson) channel. In more general versions of the model, interactions in 
the vector and axial-vector channels are treated in the same way.

In the treatment of the model considered here, the nucleon is described by 
a three-quark vertex function which satisfies a four-dimensional Faddeev
equation. The local nature of the quark-quark interaction means that the 
structure of this integral equation is similar to that of a BS equation. 
At lowest order the kernel $K$ can be pictured as exchange of 
a quark between a quark and two interacting quarks (a ``diquark"). 

The coupling strengths of the quark-quark interaction can be found by
Fierz-transforming it into these channels\cite{IBY}. The colour-$\bar 3$ diquark
channels are the relevant ones for formation of a nucleon. For a local
interaction, there are two such channels: scalar isoscalar and axial-vector
isovector. The transformed interaction Lagrangian has the form
\be
{\cal L}_I=g_{\scriptscriptstyle S} 
\sum_a\left(\bar{\psi}\gamma_5 C\tau_2\beta^a\bar{\psi}^T\right)
\left(\psi^T C^{-1}\gamma_5\tau_2\beta^a\psi\right)
+g_{\scriptscriptstyle A} 
\sum_{i,a}\left(\bar{\psi}\gamma_\mu C\tau_i\tau_2\beta^a\bar{\psi}^T
\right)\left(\psi^T C^{-1}\gamma^\mu\tau_2\tau_i\beta^a\psi\right)
\ee
where
\be
(\beta^a)_{ik}=i\sqrt{\frac{3}{2}}\epsilon_{aik}, 
\ee
and $g_s$ and $g_a$ are the couplings in these channels.\footnote{Other 
contributions including colour-6 diquark channels and the chiral partner of the 
axial diquark, $\bar{\psi}\gamma_\mu\gamma_5 C\tau_2\beta^a\bar{\psi}^T$, are
generated by the Fierz-transformation. We have not written these down since
they are not needed for the present discussion.} For the original NJL 
model, these couplings can be related to $g_\pi$ as described in Ref.\cite{IBY}.
In the scalar-diquark channel, the propagation of two interacting quarks can be 
described by the $T$-matrix,
\be
D_s(p)=\frac{4ig_s}{1-2ig_s\Pi_s(p^2)}, \eqn{D}
\ee
where $\Pi_s(p)$ has the same form as in the pion channel, \eq{Pip}. The 
interaction between two quarks in the axial diquark channel leads to a $T$-matrix 
whose form is similar to that in a vector-meson channel (see 
Ref.\cite{IBY} for more details).

The corresponding Faddeev equation is represented diagrammatically in \fig{bs}.
It can be written in the form of \eq{G} by defining the ``free" quark-diquark
propagator
\be
G_0(P)=S(P-p)D_s(p).
\ee
The quark-diquark interaction kernel is given by the quark exchange
term
\be
K_{a'a}(p',p)=\gamma_5\beta^a S(p'+p)\gamma_5\beta^{a'}.  \eqn{Z}
\ee
When the axial diquark channel is included, the kernel must be extended to a
$2\times 2$ matrix which couples the two spin-isospin channels, as
described in Ref.~\cite{IBY}.

\begin{figure}[t]
\centerline{\epsfxsize=9cm\epsfbox{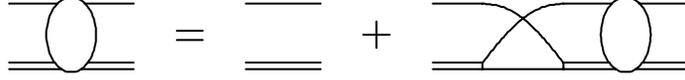}}
\vspace{3mm}

\caption{\fign{bs} The Faddeev equation for the quark-diquark Green function.}
\end{figure}

\section{Mesonic corrections to the nucleon}

The mesonic corrections to this model can be found by applying the 
general formulation of \eqs{G}--(\ref{Delta}) to the Faddeev equation outlined
above. The resulting expression for $\Delta$ is represented diagrammatically 
in \fig{fig:Delta}. Mesonic corrections to the nucleon in the NJL model have 
previously been studied by Ishii\cite{ishii} but it should be noted that the 
set of diagrams in that work is incomplete. Only the diagrams of \fig{ishii} 
were included, with subtracted meson propagators as discussed below.

\begin{figure}[t]
\centerline{\epsfxsize=16cm\epsfbox{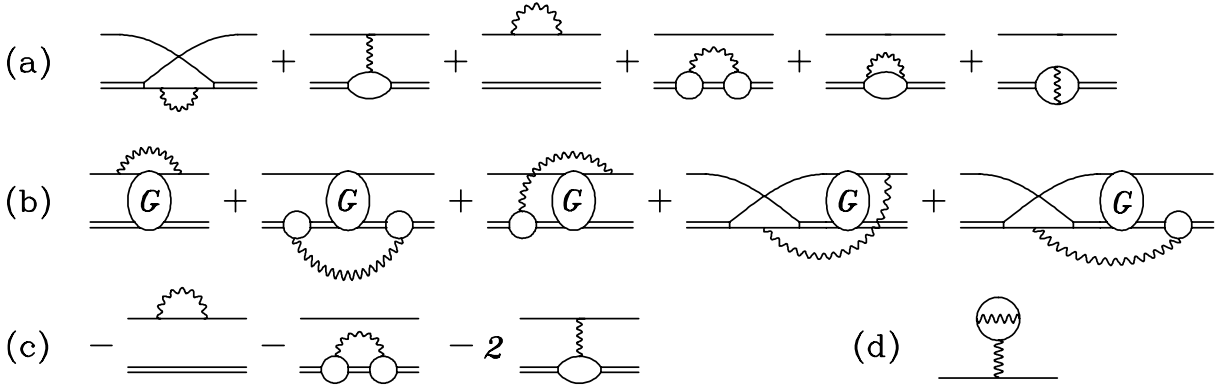}}
\vspace{4mm}

\caption{\fign{fig:Delta} The mesonic corrections, $\Delta$,
to the nucleon in the NJL model. Wavy lines denote meson propagators 
in the RPA and double lines the corresponding diquark propagators. 
In line (a) the first diagram is the term $\delta K$ in \eq{Delta}, 
and the rest form $G_0^{-1}\delta G_0 G_0^{-1}$. The diagrams of line 
(b), along with the ones formed by swapping initial and final states 
in the last three diagrams, make up $\Gamma^\mu G\Gamma^\nu D_{\mu\nu}$.
Those of line (c) form the subtraction term 
$-\Gamma_0^\mu G_0\Gamma_0^\nu D_{\mu\nu}$. In addition tadpole-type
insertions on the quark lines need to be included. For reasons of 
space, we show only one representative diagram of this class as (d).}
\end{figure}

The full set of diagrams includes not only the meson exchanges between the 
quarks treated by Ishii\cite{ishii} but also the meson clouds of the quark
and diquark. In calculating the pion-cloud contributions it is essential to
attach the pion in all possible ways to the dressed quark propagator 
obtained from the Hartree approximation\cite{DSTL,PB}. As well as direct 
dressings of the quark lines, as in \fig{fig:Delta}(a), this includes 
attachments within the Hartree self-energy and generates tadpole diagrams 
like \fig{fig:Delta}(d) where a pion loop is connected to a quark line via 
an interacting quark-antiquark pair in the scalar, isoscalar channel. The 
exchanged quark-antiquark pair is represented in \fig{fig:Delta}(d) by a 
sigma-meson propagator. There is a strong cancellation between these 
diagrams and the basic pion loop attached to a single quark. This 
cancellation is needed to get the correct leading nonanalytic (LNA) terms 
in the chiral expansion in any model where the pion-fermion coupling has a 
pseudoscalar ($\gamma_5$) form\cite{DSTL,NBCRG,oertel,PB}.

The complete one-meson correction to the nucleon mass in this model can be 
calculated as the expectation value of $\Delta$ in the lowest-order wave function 
obtained from the homogeneous equation corresponding to \eq{G}. This can be written
\be
\delta M=\overline\Psi\Delta\Psi,
\ee
where $\Psi$ satisfies
\be
\Psi=G_0K\Psi.
\ee

In models where the mesons are not introduced as elementary fields,
some care should be taken with the quark-antiquark $T$-matrix, 
\eq{Dp}. In the mesonic dressing of a single quark or one-loop 
corrections to meson properties, this $T$-matrix can be used as a meson 
propagator to generate a complete set of terms at next-to-leading order in
$1/N_c$ where $N_c$ is the number of colours\cite{oertel,ripka,PB}. 
These include terms which 
are linear in the coupling $g_\pi$, such as the 
exchange or Fock term 
corresponding to the Hartree self-energy of \eq{Hartree}. 
For a diquark or a baryon, these linear terms are already included in the 
interactions between the quarks. Hence meson exchanges between two
quarks which do not otherwise interact should be described by the subtracted
$T$-matrix
\be
\overline D_{\mu\nu}(p)=D_{\mu\nu}(p)-4ig_\pi\delta_{\mu\nu}. \eqn{Dsub}
\ee
In the case of the pion, this has the form
\be
\overline D_\pi=\frac{-8g^2_\pi\Pi_\pi(p^2)}{1-2ig_\pi\Pi_\pi(p^2)}.
\ee
This subtracted $T$-matrix should be used in the diagrams of \fig{ishii}, 
which form part of \fig{fig:Delta}.
\begin{figure}[t]
\centerline{\epsfxsize=16cm\epsfbox{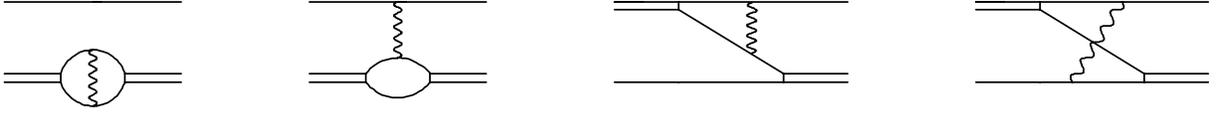}}
\vspace{4mm}

\caption{\fign{ishii} The parts of the mesonic corrections
 where the 
subtracted meson propagator of \eq{Dsub} should be used.}
\end{figure}

For a single quark (or a meson) the mesonic one-loop diagrams, 
including the Fock terms, form the leading corrections to the properties
of these particles in a systematic expansion in powers of $1/N_c$.
In the case of baryons, things are not so clean. Both Hartree and Fock 
terms contribute to the baryon energy at leading order ($N_c^1$).
Meson exchanges between the quarks in a baryon are also of leading order, 
as are nonlinear terms arising from interactions among the mesons. In
the large-$N_c$ limit, this provides a justification for treating the
baryon as a soliton, as in the Skyrme model. Such treatments have been
quite widely applied to NJL-type models\cite{alkofer,christov}. 
They form a radically different approach to baryon structure from the 
one outlined here and it 
is still unclear which is the better for the case of interest, $N_c=3$. 
Indeed a hybrid of the two has been proposed in Ref.\ \cite{zueck}.
Ultimately, numerical calculations of the size of the kernel based on
\eq{Dsub} will indicate whether a perturbative treatment is adequate.

If it is found that the shorter-range pieces of $\overline D$ are too
strong to be treated perturbatively, then one possible way to improve the
convergence of the approach may be to vary the strength of the 
subtraction term. This must be compensated by adjusting the strength of 
the interaction used in the bare equation for $G$. That would allow
the strong short-range effects to be treated self-consistently in
this equation, leaving the longer-range effects to act as a residual
interaction.

Note that the loop integral \eq{Pip} falls off at high momenta in
this type of model. Consequently the subtracted $T$-matrix \eq{Dsub} 
also falls off; the enhancements due to the Goldstone-boson nature of
the pion only appear at low momenta. It is thus the underlying 
interaction responsible for spontaneous symmetry breaking which controls 
the short-range interaction between quarks, not the short-range part of 
Goldstone-boson-exchange as suggested by some authors\cite{GR}. 
Provided that the quarks in a baryon can be assumed to interact in 
$s$-waves, only the strengths of the interactions in the scalar and
axial diquark channels are relevant.

The inclusion of interactions between the quarks while a meson is in 
flight, \fig{fig:Delta}(b), is crucial to ensuring the correct energy 
denominators for the intermediate baryon-meson states. The binding
energy of the nucleon depends on both the scalar and axial diquark
couplings, whereas that of the $\Delta$ depends only on the axial one.
The short-range interaction thus has an implicit spin-dependence, which 
splits the bare nucleon and $\Delta$ states. This means that, provided 
the scalar and axial diquark couplings are chosen to give the correct 
N-$\Delta$ splitting, the only intermediate states in the diagrams of
\fig{fig:Delta}(b) with energy denominators of order $m_\pi$ are 
pion-nucleon ones. These will generate the correct LNA terms in a chiral 
expansion, for example the order-$m_\pi^3$ term in the nucleon 
mass\cite{BKM}.

A model of this type can therefore help to resolve some of the issues which have
arisen in comparisons of the results of chiral quark models with those of the
cloudy bag model \cite{TK1,TK2,gloz1,gloz2}.  In order to get the correct LNA
terms (which are a signal that chiral symmetry has been correctly implemented)
it is crucial that the full propagator $G$ is used, allowing the quarks to
interact while the meson is in flight. This means that the nucleon-plus-pion
states necessary for the LNA terms are included, as in the cloudy bag model
\cite{thomas}. However, in contrast to the elementary bag model (i.e.\ one
defined strictly by the cloudy bag model Lagrangian \cite{thomas}), in our
approach the spin-dependence of the interaction binding the ``bare'' baryons
means that the $\Delta$ is not degenerate with the nucleon and so it does not
contribute to the LNA terms in the nucleon mass.  Although this problem is
avoided in practical bag model calculations where the bare baryon in the
baryon-plus-pion state is replaced by a physical one, such an approach is,
apparently, equivalent to resumming a subset of mesonic corrections to all orders
in the Lagrangian-based description. Such a sum may give rise to conflicts with
chiral symmetry at higher orders. In a related context, it has been noted that if
$\pi N$ loops are resummed in this way then one should also dress the
$\pi N$ vertex with pion loops\cite{OT}.

The chiral quark model of Glozman and coworkers \cite{GR,gloz1,GR2}
includes pion loops which dress the quarks individually, effectively treating
the quarks as non-interacting while the pion is in flight. In addition, 
retardation is neglected in the pion exchanges between quarks. These two 
approximations mean that the current treatments of this model do not generate 
the correct LNA terms. These would require the inclusion of interactions
between the quarks while the meson is in flight, together with the corresponding
retardation effects.

Finally it is also worth noting that, in a model of the NJL type, a complete 
set of intermediate baryon states can be included in the meson-loop diagrams. 
This will make it possible to examine the role of excited baryons in the 
mesonic dressing. Although the contributions of individual excited states 
fall off with energy, there is an infinite number of such states, which may 
be unbound as well as bound in a nonconfining model. The sum over these
states would diverge in the absence of a cut-off. We therefore expect
that they could make significant contributions to the mesonic 
dressing in the approach described here. In this way it will be possible to 
assess the accuracy of truncating to intermediate nucleon and $\Delta$ states 
only, as done in the original cloudy bag model\cite{thomas} and more recently 
in a model of the type discussed here\cite{HORSTT,OT}.

In this recent work, Oettel and coworkers\cite{HORSTT,OT} have calculated 
pion-loop contributions to nucleon properties within a Faddeev approach, with a
covariant version of the approximation used in the cloudy-bag model. In the 
original cloudy-bag model\cite{thomas}, pion loops were calculated keeping 
only $\pi N$ and $\pi\Delta$ intermediate states and the baryons were 
treated as static. In Refs.\ \cite{HORSTT,OT} these loops are calculated using
fully covariant propagators for the nucleon and $\Delta$ but keeping only the 
poles corresponding to the lowest-energy solutions to the Faddeev equation in 
each channel. In esssence, this replaces their relativistic quark model by a 
model of elementary baryons and mesons. Oettel {\it et al}.\ also take care
to include the tadpole diagrams required for the correct chiral behaviour.
Their loop integrals are made finite by
form factors at the meson-baryon vertices. These form factors can be regarded 
either as {\it ad hoc} regulators of the sort typically used in NJL models or 
as consequences of the internal structure of the hadrons. Although this approach 
respects chiral constraints and provides good estimates of the largest individual
loop contributions, it does not give the complete loop correction since excited 
baryon states have been omitted.

\section{Mesonic dressing of currents}

In order to calculate electromagnetic and weak properties of the nucleon, we
need to construct the corresponding currents. At lowest order (no meson loops)
the electromagnetic current operator to be used in a three-quark state can be 
obtained diagramatically by attaching a photon everywhere in \eq{G}. The 
resulting expression for the matrix element of the current has the form
\be
J^\mu=\bar\Psi(\tilde\Gamma_0^\mu + \tilde\K^\mu)\Psi.   \eqn{LOcur}
\ee
Here $\tilde\Gamma_0^\mu$ denotes the sum of single-quark and diquark currents,
and $\tilde\K^\mu$ the interaction current where the photon is attached to 
the exchanged quark. This corresponds to the impulse approximation in the 
three-quark basis, properly subtracted to avoid double counting \cite{bk}.
The corresponding axial current can be found in the same way by making an 
axial-vector insertion instead of attaching a photon. Currents like 
\eq{LOcur} have been extensively studied in the NJL model \cite{njl} 
as well as in QCD-motivated Faddeev \cite{oettel}
and SD approaches \cite{roberts}.
 The same methods can also be 
used to calculate the matrix element of any operator which is bilinear in
the quark fields, such as the pion-nucleon sigma commutator.

At the next order, we need to consider the one-loop mesonic contributions
to the currents, $\tilde\Delta^\mu$. These have a similar origin to the
contributions to the kernel, $\Delta$, and so can be obtained by attaching
a photon everywhere in $\Delta$, including the meson. The attachments to the 
composite mesons and diquarks are realised by direct couplings to quark 
currents in the $q\bar q$ and $qq$ loops respectively.

In addition we need to take into account the effects of mesonic fluctuations 
on the bound state wave function. If the full wave function is written as
$\Psi+\delta\Psi$, it satisfies the Bethe-Salpeter-type equation
\be
(\Psi+\delta\Psi)=G_0(K+\Delta)(\Psi+\delta\Psi).
\ee
The perturbation of the wave function $\delta\Psi$ can be expressed 
as\cite{kvin}
\be
\delta\Psi=\left[G_B\Delta-\frac{1}{2M}\left(\bar\Psi P^\mu
\frac{\partial\Delta}{\partial P^\mu}\Psi\right)_{|P|=M}\right]\Psi.
\ee
In the first term, $G_B$ denotes the unperturbed Green function $G$ with the 
bound-state pole subtracted off. This term incorporates configuration-mixing 
effects generated by meson-exchange between the quarks. The second term in 
this expression is just a wave-function renormalisation, analogous to that 
in the cloudy bag model. It involves the derivative of the meson-exchange
part of the kernel with respect to the the total 4-momentum, evaluated at the 
unperturbed nucleon mass. 

The full one-loop contribution to the current matrix element is finally
\be
\delta J^\mu=\bar\Psi\tilde\Delta^\mu\Psi+
\bar\Psi(\tilde\Gamma_0^\mu + \tilde\K^\mu)\delta\Psi+
\delta\bar\Psi(\tilde\Gamma_0^\mu + \tilde\K^\mu)\Psi. \eqn{NLOcur}
\ee

Again, it is important that $\tilde\Delta^\mu$ contain the full propagator 
$G$, allowing the quarks to interact while the meson is in flight, see 
\fig{fG}. 
\begin{figure}[t]
\centerline{\epsfxsize=10cm\epsfbox{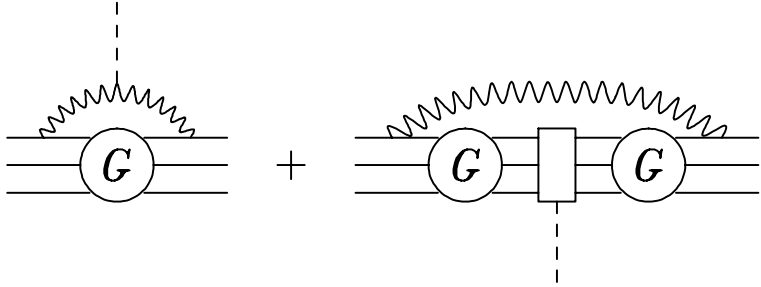}}
\vspace{4mm}

\caption{\fign{fG} Mesonic corrections of three-quark (baryon) currents
where the quarks interact while the meson is in flight. The dashed line
denotes an external photon or other operator.} 
\end{figure}
As in the case of the nucleon mass, this generates the correct LNA terms in 
chiral 
expansions of the electromagnetic radii of the nucleon\cite{BKM}. 
In particular the isovector charge radius contains a term of 
order $\ln m_\pi$ and the isovector magnetic radius one of order $m_\pi^{-1}$.
In both cases these LNA terms contribute significantly to the full radii and
so any calculation should include pion-nucleon loops with the correct energy 
denominators.

Neglecting the interactions while the pion is in flight would lead to 
expressions for the currents like those in the quark model of Glozman and 
coworkers\cite{GR,GR2,WBKPR}, which do not contain the correct LNA terms.
Truncating the sum over intermediate states to the nucleon and $\Delta$ only
would lead to an approximation similar to the cloudy-bag model, with the 
correct LNA terms\cite{TK2}.

\section{Conclusion}

We have presented a method for constructing a complete set of one-loop mesonic  
corrections to the nucleon in a chiral quark model, treated in a covariant manner. 
Our approach is based on the ``gauging of equations" method which has previously 
been used to construct electromagnetic corrections to hadronic 
systems. It ensures that a complete set of one-loop corrections is generated, 
with no double counting. This completeness is essential if physical quantities 
calculated in this approach are to satisfy the constraints of chiral symmetry
and gauge invariance. 

To illustrate the method, we have discussed in some detail its application to a 
particular covariant chiral quark model: the NJL model. The method starts from
an unperturbed valence-quark Green's function, which in this case is obtained
from a Faddeev equation. The complete set of mesonic insertions is used to
generate the one-loop contribution to the kernel. The same method can
also be used to construct photon-nucleon couplings in a way that respects
current conservation.

A central feature of this approach is the use of the unperturbed Green's function
for three interacting valence quarks. This means that while a meson is in flight
the quarks propagate in physical baryon states, as described at leading order
in the meson-loop expansion. The fact that the NJL model gives rise to an 
N-$\Delta$ splitting at this order means that the only nonanalytic terms in the 
chiral expansion are generated by pion-nucleon states. This ensures that the 
method will give the correct leading nonanalytic terms in nucleon observables, in
agreement with ChPT. This is particularly important for electromagnetic radii, 
where these nonanalytic terms make significant contributions.

The usual treatment of the cloudy-bag model similarly gives the correct leading 
nonanalytic terms in nucleon observables. However in models of the type discussed
here, there is no need to truncate the unperturbed Green's function to just
nucleon and $\Delta$ states. A complete set of baryon states can be summed over, 
regulated by the cut-off which forms an intrinsic part of any NJL-type model.

In contrast, the chiral quark model of Glozman and coworkers effectively
uses a free Green's 
function to describe the propagation of the quarks in 
meson-baryon intermediate
 states. It also neglects retardation effects on the
exchange of pions between quarks. As a result this model, in the framework used so 
far, fails to give the 
correct nonanalytic terms in the chiral expansion. However 
this could be remedied by
 a treatment based on the approach described here, or the 
corresponding time-ordered
 version.

\section*{Acknowledgements}
This work was supported by the EPSRC. We are grateful to T. Cohen
and L. Glozman for helpful discussions.


\begin{thebibliography}{99}
\bibitem{BKM} V. Bernard, N. Kaiser and U.-G. Meissner, Int. J. Mod. Phys. 
{\bf E4} (1995) 193 [hep-ph/9501384].
\bibitem{thomas} A. W. Thomas, Adv. Nucl. Phys. {\bf 13} (1984) 1.
\bibitem{birse} M. C. Birse, Prog. Part. Nucl. Phys. {\bf 25} (1990) 1.
\bibitem{CCQM} K. Shimizu, Phys. Lett. {\bf B148} (1984) 418; 
D. Robson, Phys. Rev. {\bf D35} (1985) 1029; 
K. Maltman, Nucl. Phys. {\bf A446} (1985) 623.
\bibitem{fern} F.Fernandez, A. Valcarce, U. Straub and A. Faessler, 
J. Phys. G:  Nucl. Part. Phys. {\bf 19} (1993) 2013.
\bibitem{GR} L. Ya. Glozman and D. O. Riska, Phys. Rep. {\bf 268} (1996) 263
[hep-ph/9505422].
\bibitem{GPPVW}L. Ya. Glozman, Z. Papp, W. Plessas, K. Varga and R. F. Wagenbrun,
Phys. Rev. {\bf C57} (1998) 3406 [nucl-th/9705011].
\bibitem{klev} S. P. Klevansky, Rev. Mod. Phys. {\bf 64} (1992) 649.
\bibitem{buck} A. Buck, R. Alkofer and H. Reinhardt, Phys. Lett. B {\bf 286} (1992) 
29.
\bibitem{IBY} N. Ishii, W. Bentz and K. Yazaki, Phys. Lett. B {\bf 318} (1993) 26; 
Nucl. Phys. {\bf A587} (1995) 617. 
\bibitem{huang} S. Huang and J. Tjon, Phys. Rev. C {\bf 49} (1994) 1702.
\bibitem{alkofer} R. Alkofer, H. Reinhardt and H. Weigel, Phys. Rep. 
{\bf 265} (1996) 139 [hep-ph/9501213].
\bibitem{christov} C. V. Christov et al., Prog. Part. Nucl. Phys. {\bf 37} 
(1996) 91 [hep-ph/9604441].
\bibitem{zueck} U. Z\"uckert, R. Alkofer, H. Weigel and H. Reinhardt,
Phys. Rev. {\bf C55} (1997) 2030 [nucl-th/9609012].
\bibitem{ishii} N. Ishii, Phys. Lett. {\bf B431} (1998) 1.
\bibitem{HORSTT}M. B. Hecht, M. Oettel, C. D. Roberts, S. M. Schmidt, 
P. C. Tandy and A. W. Thomas, Phys. Rev. {\bf C65} (2002) 055204
[nucl-th/0201084].
\bibitem{OT}M. Oettel and A. W. Thomas, nucl-th/0203073.
\bibitem{BKR}P. J. A. Bicudo, G. Krein and J. E. F. T. Ribeiro,  Phys. Rev. 
{\bf C64} (2001) 025202 [hep-ph/0105289].
\bibitem{nnn} A. Kvinikhidze and B. Blankleider, Phys. Rev. {\bf C60} (1999) 
044003, 044004 [nucl-th/9901001, nucl-th/9901002].
\bibitem{Adel98} A.\ Kvinikhidze and B.\ Blankleider, in {\it Nonperturbative 
Methods in Quantum Field Theory}, edited by A. W. Schreiber, A. G. Williams,
and A. W. Thomas (World Scientific, Singapore, 1998), p.284 [nucl-th/9806046];
Nucl. Phys. {\bf A670} (2000) 210c [nucl-th/9906017].
\bibitem{DSTL}V. Dmitra\u{s}inovi\'{c}, H.-J. Schulze, R. Tegen and R. H.
Lemmer, Ann. Phys. (N.Y.) {\bf 238} (1995) 332.
\bibitem{NBCRG}E. Nikolov, W. Broniowski, C. V. Christov, G. Ripka and K. Goeke,
Nucl. Phys. {\bf A608} (1996) 411.
\bibitem{oertel} M. Oertel, M. Buballa and J. Wambach, Nucl. Phys. {\bf A676} 
(2000) 247 [hep-ph/0001239].
\bibitem{PB} R. S. Plant and M. C. Birse, Nucl.Phys. {\bf A703} (2002) 717 
[hep-ph/0007340].
\bibitem{TK1} A. W. Thomas and G. Krein, Phys. Lett. {\bf B456} (1999) 5 
[nucl-th/9902013].
\bibitem{TK2} A. W. Thomas and G. Krein, Phys. Lett. {\bf B481} (2000) 21 
[nucl-th/0004008].
\bibitem{gloz1} L. Ya. Glozman, Phys. Lett. {\bf B459} (1999) 589 [hep-ph/9904459].
\bibitem{gloz2} L. Ya. Glozman, Phys. Lett. {\bf B494} (2000) 58 [hep-ph/0004229].
\bibitem{ripka} G. Ripka,  Nucl. Phys. {\bf A683} (2001) 463 [hep-ph/0003201].
\bibitem{GR2} L. Ya. Glozman and D. O. Riska, Phys. Lett. {\bf B459} (1999) 49
[hep-ph/9812224].
\bibitem{bk} B. Blankleider and A.N. Kvinikhidze, Phys. Rev. {\bf C62} 039801 (2000)
[nucl-th/9912003].
\bibitem{njl} H. Asami, N. Ishii, W. Bentz and K. Yazaki, Phys. Rev. {\bf C51}
(1995) 3388.
\bibitem{oettel} M. Oettel, R. Alkofer and L. von Smekal, Eur. Phys. J. {\bf A8}
(2000) 553 [nucl-th/0006082].
\bibitem{roberts} J. C. R. Bloch, C. D. Roberts and S. M. Schmidt, Phys. Rev. 
{\bf C61} (2000) 065207 [nucl-th/9911068].
\bibitem{kvin} A. N. Kvinikhidze and B. Blankleider, hep-th/0104053
\bibitem{WBKPR} R. F. Wagenbrunn, S. Boffi, W. Klink, W. Plessas and M. Radici,
Phys. Lett. {\bf B511} (2001) 33 [nucl-th/0010048].
\end{thebibliography}
\end{document}